%% file: main.tex
\documentclass[sigplan]{acmart}

\usepackage{CJKutf8}
\usepackage{graphicx}
\usepackage{amsfonts}
\usepackage{amsmath,bm}
\usepackage{url}
\usepackage{endnotes}
\usepackage{algorithm}
\usepackage{algorithmic}
\usepackage{booktabs}
\usepackage{color}
\usepackage{subfigure}
\usepackage{tabularx}
\usepackage{multirow}
\usepackage{verbatim}
\usepackage{pifont}
\newcommand{\ra}[1]{\renewcommand{\arraystretch}{#1}}

\usepackage{xcolor}

\newcommand{\ie}{\emph{i.e.,~}}
\newcommand{\etal}{\emph{et al.~}}

\newcommand{\argmax}{\operatornamewithlimits{argmax}}

\newcommand{\PreserveBackslash}[1]{\let\temp=\\#1\let\\=\temp}
\newcolumntype{C}[1]{>{\PreserveBackslash\centering}p{#1}}
\newcolumntype{R}[1]{>{\PreserveBackslash\raggedleft}p{#1}}
\newcolumntype{L}[1]{>{\PreserveBackslash\raggedright}p{#1}}

\newfont{\mycrnotice}{ptmr8t at 7pt}
\newfont{\myconfname}{ptmri8t at 7pt}

\begin{document}
\begin{CJK*}{UTF8}{gbsn}
%\thispagestyle{empty} \maketitle \thispagestyle{empty}

%\copyrightyear{2018}
\acmYear{2019}

\title{iCap: Interactive Image Captioning with Predictive Text}

\author{Zhengxiong Jia and Xirong Li}
\affiliation{Renmin University of China}

\begin{abstract}
In this paper we study a brand new topic of interactive image captioning with human in the loop. Different from automated image captioning where a given test image is the sole input in the inference stage, we have access to both the test image and a sequence of (incomplete) user-input sentences in the interactive scenario. We formulate the problem as \emph{Visually Conditioned Sentence Completion (VCSC)}. For VCSC, we propose asynchronous bidirectional decoding for image caption completion (ABD-Cap). With ABD-Cap as the core module, we build iCap, a web-based interactive image captioning system capable of predicting new text with respect live input from a user. A number of experiments covering both automated evaluations and real user studies show the viability of our proposals.
\end{abstract}


\maketitle

%===========================================================================
\section{Introduction} \label{sec:intro}
%===========================================================================

\input{intro}

%===========================================================================
\section{Related Work} \label{sec:related}
%===========================================================================
\input{related}

%The proposed ABD-CAP framework differs from previous work that attempt to exploit bidirectional modelling for image captioning in two dimensions: first, instead of capturing bidirectional visual-language interactions to mitigate the issue of unbalanced outputs, we  

%===========================================================================
\section{Human-in-the-loop Image Captioning} \label{sec:method}
%===========================================================================

\input{method}

%===========================================================================
\section{Evaluation} \label{sec:exp}
%===========================================================================

\input{exp}
\section{Conclusions} \label{sec:conc}
%===========================================================================

In this paper we have made a novel attempt towards image captioning with human in the loop. We develop iCap, an interactive image captioning system. We conduct both automated evaluations and user studies, allowing us to draw conclusions as follows. With the assistance of visually conditioned sentence completion, better image captions can be obtained with less amount of human workload. Moreover, asynchronous bidirectional decoding is found to be important for effectively modeling live input from a user during the interaction.

\bibliographystyle{ACM-Reference-Format}
\bibliography{icap}

\clearpage\end{CJK*}
\end{document}

%% file: intro.tex
Automated description of visual content, let it be image or video, is among the core themes for multimedia content analysis and retrieval. Novel visual captioning models are being actively developed \cite{huang2019attention,vidcap-ijcai19} with increasing performance reported on public benchmark datasets such as MS-COCO~\cite{chen2015mscoco} and MSR-VTT~\cite{msrvtt}. Nonetheless, the success of the state-of-the-art models largely depends on the availability of many well-annotated training examples in a specific domain. It is apparent that an image captioning model learned from MS-COCO does not work for medical images, while a model learned from annotations with respect to human activities is ineffective for describing general images~\cite{coco-cn}. It is also recognized that one cannot directly use a dataset described in one language to build models for another language~\cite{icmr2016-chisent}. Therefore, in the days to come,  when one wants to build an effective image captioning model for a domain where annotations are in short supply, manual annotation remains indispensable.

%Thanks to the rapid development of deep learning methods, especially recent progress in visual perception \cite{faster-rcnn} \cite{resnet} and sequential generation \cite{nmt} \cite{abd-nmt} \cite{seq2seq-att} \cite{transformer} tasks, extensive progress have been made in the domain of automatic image captioning \cite{show-tell} \cite{show-att-tell} \cite{top-down} \cite{self-crit}. These methods normally employ an image representation given by one or more layer of a CNN or by a time-varying context vector extracted with an attention mechanism over spatial regions, which are decoded with a RNN based language model step by step from left to right using a greedy decoding strategy \cite{show-tell}, or more sophisticated techniques like beam search and its variants \cite{guide-vocab}.

\begin{figure}[tbh!]
    \centering
    \includegraphics[width=\columnwidth]{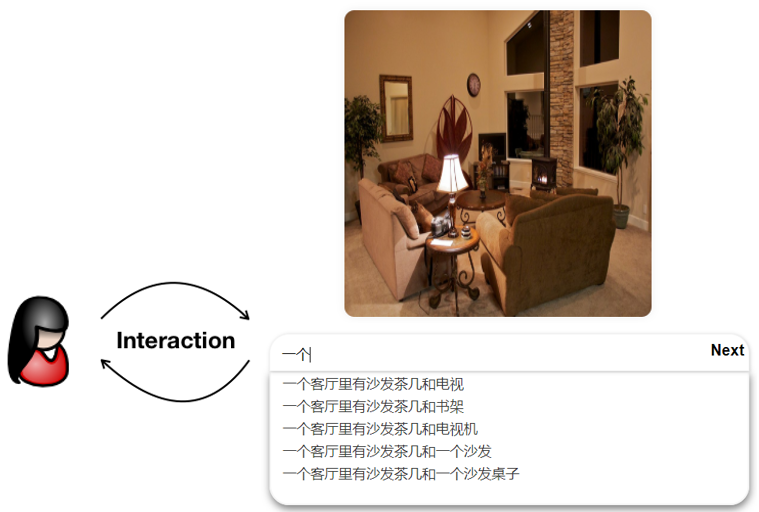}
    \caption{\textbf{User interface of the proposed interactive image captioning system}. Given both the image and live input from a user, the system interacts with the user by predicting new text to complete the user input. Note that while we target at \emph{Chinese}, which is the authors' first language,  major conclusions of this research are language-independent.}
    \label{fig:icap-system}
\end{figure}

Writing image descriptions is known to be laborious, even for crowd sourcing. Thus an important research question arises: \emph{Can manual annotation be performed in a more intelligent manner other than fully manual}? Li \etal~\cite{coco-cn} develop a recommendation assisted annotation system, where a user is presented with five sentences automatically recommended by the system based on the pictorial content. Cornia \etal \cite{show-control-tell} propose a framework that allows a user to supervise the caption generation process by specifying a set of detected objects as the control signal. Neither of them is \emph{interactive}.

In this paper, we study a novel topic of \emph{interactive image captioning} with human in the loop. Recall that in automated image captioning, a given test image is the sole input in the inference stage. By contrast, an interactive system has to consider live input from a specific user and respond on the fly, see Fig. \ref{fig:icap-system}. Thus, the interactive scenario is more challenging. A desirable system shall be capable of predicting new text that refines or completes the current text provided by the user.  Note that such a task conceptually resembles to some extent query auto completion (QAC)~\cite{review-qac} in information retrieval, as both perform text completion. Major difference is two-fold. First, QAC has no visual input. Second, different from search queries that are typically keywords or phrases, image captions are natural language text that contains richer contextual information. We term the new task \emph{Visually Conditioned Sentence Completion} (VCSC). To summarize, our contributions are as follows:
%This paper makes four contributions as follows,
\begin{itemize}
    \item To the best of our knowledge, this is the first work on interactive image captioning, where humans and deep captioning models interact with each other to accomplish the captioning task.
    \item We identify a key task in the interactive scenario, namely Visually Conditioned Sentence Completion (VCSC). We tackle the task by proposing asynchronous bidirectional decoding for image caption completion (ABD-Cap).   
    \item We verify our proposal by developing iCap, a web-based interactive annotation system that responds to user input in real time. Both automated evaluations and real user studies justify the effectiveness of the proposed ABD-Cap model and iCap system.
\end{itemize}

 %Having Drawn inspiration from the user interface that are commonly adopted by Query Auto Completion (QAC) \cite{review-qac} \cite{qc-web} in the context of information retreival, we propose an interactive captioning framework with a user interface as shown in \ref{fig:icap-system}. To realize this framework, we formulated a novel visual text generation task termed as Visually Conditioned Sentence Completion (VCSC), which differs from automated image captioning in that it takes as input both the test image and the (incomplete) user-input sentence. For VCSC, we propose asynchronous bidirectional decoding for image caption completion (ABD-Cap), with which as the core module we build iCap, a web-based interactive image captioning system capable of predicting new text with respect live input from a user. For evaluation of the proposed interactive image captioning framework, we conducted a two stage evaluation protocal with extensive experiments covering both automated evaluations and real user studies, which demonstrate the viability of our proposals. 

%% file: related.tex
\textbf{Automated image captioning}. 
A number of deep learning based methods have been proposed for automated image captioning \cite{show-tell,top-down,self-crit}. These methods mainly follows an encoding-and-decoding workflow. A given image is encoded into a dense vector by first extracting visual features by a pre-trained convolutional neural network (CNN) and then reduce the features to the dense vector either by affine transformation~\cite{show-tell} or by an spatially aware attention module~\cite{top-down}. Even though this line of work has achieved extraordinary performance on several benchmarking datasets, there are still limitations when applied in more complex scenarios especially when humans want to get control of the captioning process. 
Recent work by Cornia \etal \cite{show-control-tell} proposes a framework that allows human to control the caption generation process by specifying a set of detected objects as the control signal. Our work goes further in this direction, investigating interactive image captioning with human in the loop.
%, where full control of the captioning process is given to a user, humans while the capability of the captioning models could be fully exploited in the same time.   

\textbf{Interactive image labeling}. In the context of semantic image segmentation, several works have been done to speed up the pixel labeling process using interactive techniques. Representative works are the polygon-rnn series including polygon-rnn \cite{polygon-rnn}, polygon-rnn++ \cite{polygon-rnn++} and curve-gcn \cite{curve-gcn}, which focus on producing polygonal annotations of objects interactively with humans-in-the-loop. To the best of our knowledge, we are the first to devise an interactive annotation tool for image captioning. 
%,  of image captioning, which is also a fundamental task that relies heavily on large scale and fine-grained data annotations.  

\textbf{Bidirectional decoding}. Research on utilizing  
bidirectional decoding for image captioning exists \cite{twin-net,bi-imcap}. In \cite{twin-net}, a backward decoder is used only in the training stage to encourage the generative RNNs to plan ahead. As for \cite{bi-imcap}, a bidirectional LSTM (Bi-LSTM) combining two separate LSTM networks is developed to capture long-term visual-language interactions. Predictions at a specific time step is fully determined by the forward and backward hidden states at that step.
%Standard neural sequence decoders generate target sentences from left to right, and it has been proven to be important to establish the direct information flow between currently predicted word and previously generated words \cite{past-future-nmt} \cite{transformer}. However, most current methods still fail to estimate some desired information in the future and resulted in the issue of unbalanced outputs. To address this problem, a bunch of work exploited a bidirectional decoding scheme to capture future contexts in the generation process \cite{abd-nmt} \cite{twin-net} \cite{sync-bi-nmt} \cite{bi-imcap}. Among this work, \cite{twin-net} and \cite{bi-imcap} have ever experimented with a bidirectional decoding scheme for image captioning. To be specific, \cite{twin-net} exploited a backward decoder only in the training stage to encouraging the generative RNNs to plan ahead, and \cite{bi-imcap} exploited a deep bidirectional LSTM (Bi-LSTM) model that consists of two separate LSTM networks to capture long-term visual-language interactions, in which the prediction at current timestep are only determined by the forward and backward hidden states at this step. 
Different from Bi-LSTM, the two LSTMs used in our ABD-Cap model work in an asynchronous manner that is suited for the VCSC task. This design is inspired by ABD-NMT~\cite{abd-nmt} in the machine translation field. We adapt the backward decoding process to generate a fixed-length sequence of backward hidden vectors and change the training procedure from joint training to training two decoders sequentially. We find in preliminary experiments that  the adaption is better than the original ABD for image captioning.

%To better cope with the requirements of the proposed VCSC task, we experimented a new asynchronous bidirectional decoding scheme for image captioning that are originated from the previous abd-nmt \cite{abd-nmt}, in which the two decoders are separately trained in the training stage while cooperated together in an asynchronous manner during the inference stage.    

%% file: method.tex
\subsection{Problem Formalization} 

For writing a sentence to describe a given image $I$, a user typically conducts multiple rounds of typing and editing. In such a manual annotation session, one naturally generates a sequence of (incomplete) sentences, denoted as $\{S_i | i=1,\ldots,T\}$, where $S_i$ indicates the sentence after $i$ rounds and $T$ is the number of rounds in total. Accordingly, $S_1$ is the initial user input, while $S_T$ is the final annotation. Our goal is to devise an image captioning (iCap) framework so that the user can reach $S_T$ in fewer rounds and with reduced annotation workload. 

Given the image $I$ and the user input $S_i$, it is reasonable to assume that $S_i$ is thus far the most relevant with respect to $I$. Hence, we shall take both into account and accordingly suggest $k$ candidate sentences that complete $S_i$. Apparently, this scenario differs from automated image captioning where the image is the sole input. We term the new task as \emph{Visually Conditioned Sentence Completion} (VCSC). Image captioning can be viewed as a special case of VCSC that has no user input.

%In this research, our goal is to devise a novel interactive image captioning framework (iCap), in which human and deep captioning models are able to interact with each other in a human-in-the-loop captioning session when composing an image caption. To realize this, we propose a new breed of completion algorithmes for the task of visually-conditioned sentence completion (VCSC) in which the image captioning model is exploited to generate a list of completion candidates for a given user input based on current visual information. Even more, to tackle down the obvious defects of the unidirectional captionig model in modelling future context and better support the interact captioning scene, we designed a novel bidirectional image captioning model together with a bidirectional completion algorithm built upon it, which are better equipped for the VCSC task.

\begin{figure}[tb!]
  \centering
  \includegraphics[width=\columnwidth]{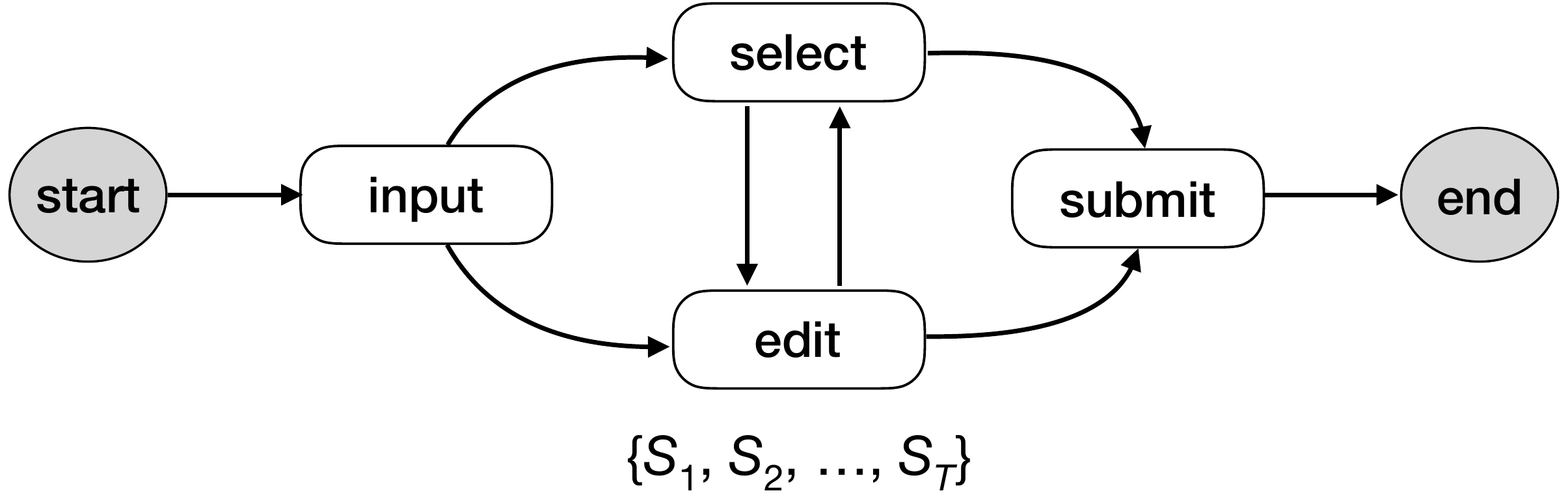}
  \caption{\textbf{User state transitions in an iCap session}. The session starts once a user types something as an initial input $S_1$. The iCap system tries to complete the input by suggesting $k$ sentences, denoted as $\{\hat{S_1}, \ldots, \hat{S_k}\}$. The user performs the captioning task by switching between the \emph{select} and \emph{edit} states, resulting in a sequence of sentences to be completed $\{S_i | i=2, 3, \ldots\}$. The session ends once the user \emph{submits} the final annotation $S_T$.}
  \label{fig:user-state}
\end{figure}

Given an iCap system equipped with a VCSC model, a user annotates an image in an iCap session, which is illustrated in Fig. \ref{fig:user-state} and described as follows.
%interacts with the system as follows. 
A session starts once the user types some initial text, \ie $S_1$. At the $i$-th round ($i\ge2$), the system presents to the user $k$ candidate sentences $\{\hat{S}_{i,1}, \hat{S}_{i,2}, \ldots, \hat{S}_{i,k}\}$ generated by the VCSC model. The user produces $S_i$ by either selecting one of the suggested sentence or revising $S_{i-1}$. The session closes when the user chooses to submit the last annotation $S_T$. 

From the above description we see that a desirable VCSC model shall not only cope with the multi-modal input but also needs to respond in real time. Next, we propose a model that fulfills these two requirements.

\subsection{Visually Conditioned Sentence Completion} \label{ssec:vcsc}

We develop our VCSC model based on Show-and-Tell~\cite{show-tell}, a classical model for automated image captioning. To make the paper more self-contained, we describe briefly how Show-and-Tell generates a sentence for a given image. Accordingly,  we explain difficulties in directly applying the model for the VCSC task.

Show-and-Tell generates a sentence by iteratively sampling words from a pre-specified vocabulary of $m$ distinct words $\{w_1,\ldots,w_m\}$. Note that in addition to normal words, the vocabulary contains three special tokens, \ie \emph{start}, \emph{end} and \emph{unk}, which indicate the beginning, the ending of a sentence and out-of-vocabulary (OOV) words. Let $p_t \in R^m$ be a probability vector, each dimension of which indicates the probability of the corresponding word to be sampled at the $t$-th iteration. Show-and-Tell obtains $p_t$ using a Long-Short Term Memory (LSTM) network~\cite{lstm97}. In particular, given $h_t \in R^d$ as the LSTM's hidden state vector, $p_t$ is obtained by feeding $h_t$ into a $d\times m$ fully connected (FC) layer followed by a softmax layer, \ie
\begin{equation} \label{eq:update-p}
\begin{array}{lll}
p_t & =& \mbox{softmax}(\mbox{FC}(h_t)), \\
\hat{w}_t & = & \argmax_{w}(p_t). 
\end{array}
\end{equation}
The word with the largest probability is sampled. Consequently, the hidden state vector is updated as 
\begin{equation} \label{eq:update-h}
h_{t+1} = \mbox{LSTM}(h_t, \hat{x}_t),
\end{equation}
where $\hat{x}_t$ is the embedding vector of $\hat{w}_t$.  To make the generation process visually conditioned,  let $v(I)$ be a visual representation of the given image extracted by a pre-trained CNN. LSTM is initialized with $h_0=\mathbf{0}$ and $\hat{x}_0$ as the visual embedding vector obtained by feeding $v(I)$ into an FC layer.

%To support the proposed interactive image captioning framework, we need to deal with a novel visual text generation scenary, which we term as visually conditioned sentence completion (VCSC). This new scenary requires a specialized sentence completion function based on current visual information, and even more, to support the interactive using case, the completion function should be able to initiate at any position of the partially completed caption where the user-specified cursor is currently pointing at.

\subsubsection{Show-and-Tell for VCSC} \label{sssec:sat-vcsc}

Assume the user input $S_i$ has $n$ words $\{w_{i,1}, \ldots, w_{i,n}\}$. Let $c$ be the user-specified cursor position, suggesting where the user wants to edit. Accordingly, $S_i$ is divided into two substrings, \ie $S_{i,l}=\{w_{i,1}, \ldots, w_{i,c}\}$ and $S_{i,r}=\{w_{i,c+1}, \ldots, w_{i,n}\}$. As the sampling process of Show-and-Tell is unidirectional, $S_{i,r}$ has to be omitted. Note that such a deficiency cannot be resolved by reversing the sampling process, as $S_{i,l}$ will then be ignored. Consequently, Show-and-Tell is unable to fully exploit the user-provided information.

In order to inject the information of $S_{i,l}$ into the LSTM network, we update the hidden state vector by Eq. \ref{eq:update-h} except that the word sampled at each step is forced to be in line with $\{w_{i,1}, \ldots, w_{i,c}\}$. Afterwards, the regular sampling process as described by Eq. \ref{eq:update-p} is applied until the \emph{end} token is selected. In order to obtain $k$ candidates, a beam search of size $k$ is performed. The best $k$ decoded beams are separately appended to $S_{i,l}$ to form $\{\hat{S}_{i,1}, \ldots, \hat{S}_{i,k}\}$.

%This brought up a novel and challenging completion task that can be formulated as follows: given an input image $I$ and a partially completed caption $S$ consisting of $m$ words $\{w_1, \ldots, w_m\}$, together with the user specified cursor position $c$, which divide $S$ into the left partial sentence $S_l = \{w_1, \ldots, w_c\}$ and the right partial sentence $S_r = \{w_{c+1}, \ldots, w_m\}$, the completion function will recommend the best $k$ completion candidates $\hat{S} = \{\hat{S_1}, \hat{S_2}, \ldots, \hat{S_k}\}$ according to the conditional probability $ P(\hat{S} |I, S_l, S_r) $.

\subsubsection{Proposed ABD-Cap for VCSC} 

\begin{figure}[tb!]
  \centering
  \includegraphics[width=\columnwidth]{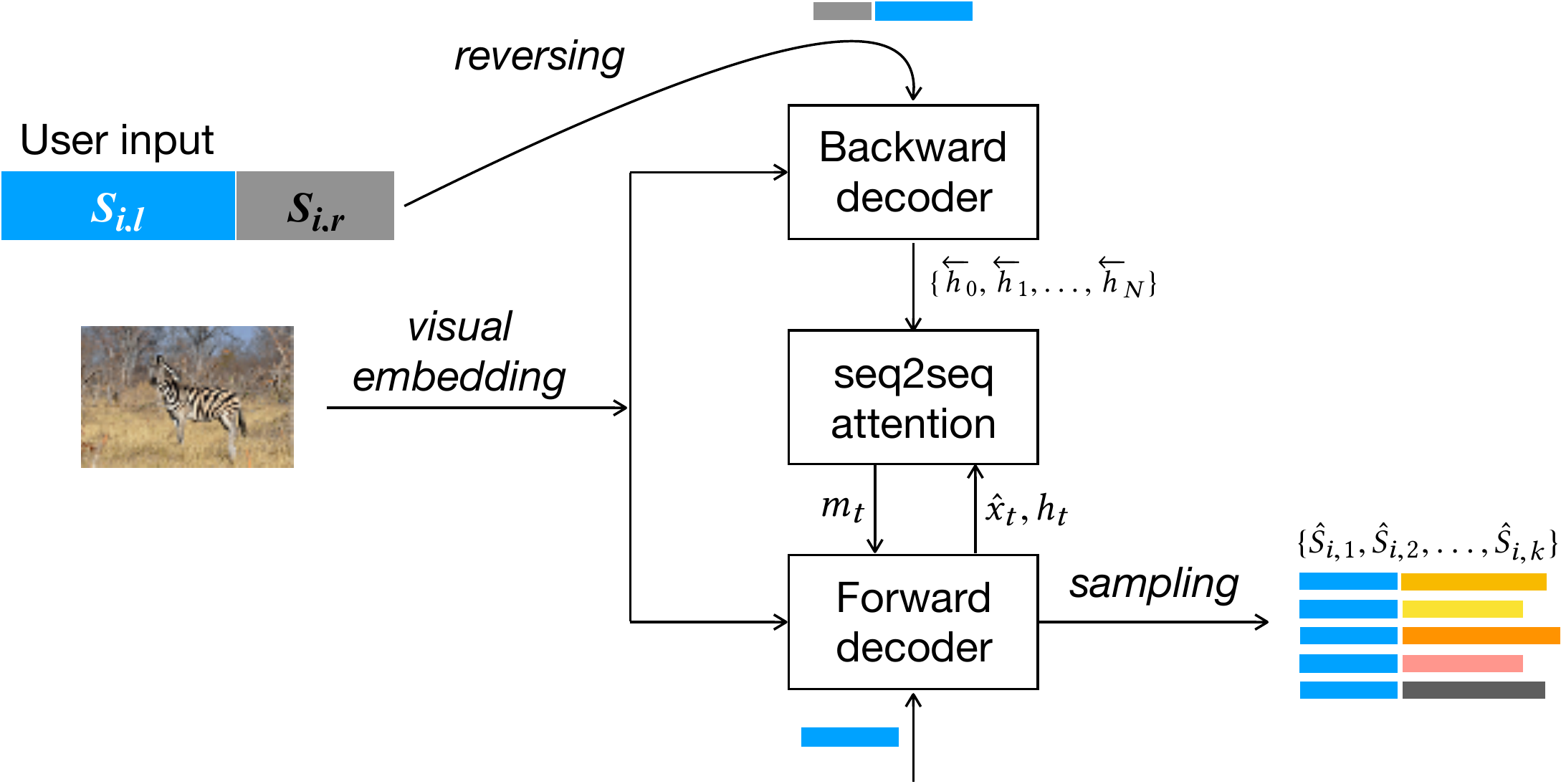}
  \caption{\textbf{Diagram of the proposed ABD-Cap model for visually conditioned sentence completion}. Given an image and a user-input sentence $S_i$ which is separated into two substrings $S_{i,l}$ and $S_{i,r}$ by the user-specified cursor, the forward encoder exploits multi-source information (from the image and the backward decoder) to complete $S_i$.}
  \label{fig:abd}
\end{figure}

In order to effectively exploit the user input on both sides of the cursor, we propose Asynchronous Bidirectional Decoding for image caption completion (ABD-Cap). The main idea is to deploy two asynchronous LSTM based decoders, see Fig \ref{fig:abd}. One decoder is to model the entire user input in a backward manner so that $S_{i,r}$ is naturally included. The other decoder is responsible for sentence generation similar to Show-and-Tell except that it receives information from the first decoder through a seq2seq attention module~\cite{seq2seq-att}. The two decoders are separately trained, and cooperate together in an asynchronous manner for sentence completion.

\textbf{Backward decoder}. Our backward decoder is trained using reserved image captions, and thus learns to generate backward hidden state vectors $\{\overleftarrow{h}_t\}$ from right to left, \ie
\begin{equation} \label{eq:update-hb}
\begin{array}{lll}
\overleftarrow{h}_{t} & = & \mbox{LSTM}(\overleftarrow{h}_{t+1}, \hat{x}_{t+1}), \\
\overleftarrow{p}_t & =& \mbox{softmax}(\mbox{FC}(\overleftarrow{h}_t)), \\
\hat{w}_t & = & \argmax_{w}(\overleftarrow{p}_t). 
\end{array}
\end{equation}

In order to cope with user input of varied length, we sample $N$ rounds, where $N$ is the maximum sequence length. This results in a fixed-length sequence of backward hidden state vectors  $\overleftarrow{H} = \{\overleftarrow{h}_{0}, \overleftarrow{h}_{1}, \ldots, \overleftarrow{h}_{N}\}$. For injecting the information of $S_i$ into the sequence, we first conduct the regular sampling procedure for $N-n$ steps. Afterwards, the word sampled per step is forced to be in line with $\{w_{i,n}, \ldots, w_{i,1}\}$.

%Given an input image $I$ and a user input sentence $S = \{w_1, \ldots, w_m\}$, we first sample with the backward decoder of the bidirectional image captioning model for $N-m$ step, where $N$ is the maximum sequence length, and from the $N-m+1$ step, we feed the reverse user input $S = \{w_1, \ldots, w_m\}$ through the backward decoder, by which process, the reverse user input sentence will be encoded into the backward hidden state sequence $\overleftarrow{H} = \{\overleftarrow{h}_{0}, \overleftarrow{h}_{1}, \ldots, \overleftarrow{h}_{N}\}$.

%The backward decoder of our model is similar to the decoder of the original Show and Tell model \cite{show-tell}, except that it's trained to generate a reverse image caption, by which procedure it also learns to generate the backward hidden state sequence from right to left.

\textbf{Forward decoder with attention}. In order to utilize the information from the backward decoder, we employ the  seq2seq attention mechanism~\cite{seq2seq-att}. In particular, $\overleftarrow{H}$ is converted to a vector $m_t$ that encodes the backward contextual information. Then, by substituting $m_t$ for $\hat{x}_t$ in Eq. \ref{eq:update-h}, the backward information is embedded into the word sampling process. We express the above as
\begin{equation}
\begin{array}{lll}
a_t &=& \mbox{softmax}(\mbox{ReLU}(\mbox{FC}(\hat{x}_t, h_t)), \\
m_t &=& \sum_{i=1}^{N} a_{t,i} \cdot \overleftarrow{h_i},\\
h_{t+1} &=& \mbox{LSTM}(h_t, m_t),
\end{array}
\end{equation}
where $a_t$ is a $N$-dimensional attention weight vector.

For the rest of the VCSC task, we follow the same procedure as depicted in Section \ref{sssec:sat-vcsc}.

\subsection{Training VCSC Models}

\input{train-models}

\medskip

For user study, we build a web-based iCap system, with its user interface shown in Fig. \ref{fig:icap-system}. Given a specific user input, it suggests $k=5$ sentences in approximately 70 milliseconds,  which is sufficiently fast for real-time interaction.

%% file: train-models.tex
Since our target language is Chinese, we train the VCSC models on the COCO-CN dataset~\cite{coco-cn}. This public dataset contains 20,342 MS-COCO images annotated with 27,218 Chinese sentences. We follow the original data split, with 18,342 images for training, 1,000 images for validation and 1,000 images for test. For image features we use the provided 2,048-dim CNN features, extracted using a pretrained ResNeXt-101 model~\cite{mettes2016imagenet}.

%For visual features, we use precomputed ResNeXt-101 features provided by ~\cite{coco-cn}.  

All models are trained in a standard supervised manner, with the cross entropy loss minimized by the Adam optimizer. The initial learning rate is set to be 0.0005. %, which decays every three epochs with a decaying factor of 0.8. 
We train for 80 epochs at the maximum. 
%We train the model for 80 maximum epochs while exploiting an early stopping mechanism with a patience of 10 epochs to prevent overfitting.
Best models are selected based on their CIDEr scores on the validation set. 
%We evaluate the model on the valiadation set after each training epoch, and the model achieveing the highest CIDEr score will be deemed as the best model for evaluation. 

Note that in the interactive scenario, a user might provide OOV words. 
To alleviate the issue, different from previous works \cite{icmr2016-chisent,coco-cn} that perform word-level sentence generation, our models compose a sentence at the character level. As shown in Section \ref{ssec:eval-auto}, this choice substantially reduces the occurrence of OOV words.

%What's different from the original paper is the preprocessing and representation of the Chinese sentence. To enable flexible on-site preprocessing of sentences, we exploit the Jieba Chinese text segmentation tool for Chinese text segmentation rather than the original BosonNLP toolkit. More importantly, to better support the visually-conditioned sentence completion function in Chinese setting, we propose to use a character-level representation when tokenizing Chinese sentences instead of the commonly used word-level representation, by which we were able to decrease the number of potential OOVs to a large extent with a relatively small vocabulary.

%\input{data-split}

%% file: exp.tex
Unlike automated image captioning, there lacks a well established evaluation protocol for interactive image captioning. We need to understand how a user interacts with the iCap system and evaluate to what extent the system assists the user. To that end, we propose a two-stage evaluation protocol as follows:
%In this section, we will realize the proposed interactive image captioning framework in Chinese setting and extensively evaluate its performance in three stages:
\begin{itemize}
    \item \textbf{Stage I}. Performed before commencing a real user study, evaluations conducted at this stage are to assess the validity of major implementation choices of the VCSC module used in the iCap system.
%    Performed   Evaluating the proposed bidirectional image captioning model for conventional image captioning task.
    \item \textbf{Stage II}. Evaluations at this stage are performed after running the iCap system for a while with adequate data collected to analyze the usability of iCap in multiple aspects.
    %  Evaluating the effectiveness of the proposed interactive image captioning session in authentic interactive annotation process.
    %\item \textbf{Stage III:} Evaluating the performance of the bidirectional completion algorithm comparing to the unidirectional completion algorithm in simulated completion scenes.
\end{itemize}

\subsection{Stage-I Evaluation} \label{ssec:eval-auto}

\input{eval-auto}

\subsection{Stage-II Evaluation} \label{ssec:eval-icap}

\input{eval-icap}

%% file: eval-auto.tex
\subsubsection{Setup}

With no user input provided, the VCSC task is equivalent to automated image captioning. So we first evaluate the proposed ABD-Cap model in the automated setting. 

\textbf{Baselines}. 
As described in Section \ref{ssec:vcsc}, compared to the Show-and-Tell model~\cite{show-tell} used in \cite{coco-cn} for Chinese captioning, we make two main changes. That is, the proposed asynchronous bidirectional decoder and character-level sentence generation. In order to verify the necessity of these two changes, we compare with the following four baselines: 
\begin{itemize}
	\item Show-and-Tell with a word-level forward decoder~\cite{coco-cn}.
	\item Show-and-Tell with a word-level backward decoder.
	\item Show-and-Tell with a character-level forward decoder.
	\item Show-and-Tell with a character-level backward decoder.	
\end{itemize}

%As the backbone of the whole framework, we first train the proposed bidirectional image captioning model and evaluate its performance for the conventional image captioning task before applying it to the interactive image captioning system.

%\textbf{Baselines.} Two dimensions of baselines are involved to justify the effectiveness of our method. The first dimension concerns the modelling methods, for which we consider two alternates: (1) the original Show and Tell model and (2) the backward Show and Tell model trained to generate a reverse image caption. The second dimension is the representation of sentences, for which we will compare the model trained with character-level sentence and word-level sentence to show that competitive performance can still be achieved with character-level representation.

For a fair comparison, the four baselines are all trained in the same setting as ABD-Cap. All the models are evaluated on the 1,000 test images of COCO-CN~\cite{coco-cn}, each of which is associated with five manually written Chinese captions.

\textbf{Evaluation criteria}. We report BLEU-4, METEOR, ROUGE-L and CIDEr, commonly used for automated image captioning. Note that computing the four metrics at the character level is not semantically meaningful. So for a sentence generated by a character-level decoder, we employ Jieba$\footnote{https://github.com/fxsjy/jieba}$, an open-source toolkit for Chinese word segmentation, to tokenize the sentence to a list of words. The presence of the \emph{unk} token in a generated sentence means an OOV word is predicted, which negatively affects user experience. Therefore, for each model we calculate the OOV rate, \ie the number of sentences containing \emph{unk} divided by the number of generated sentences.

%Note that the aformentioned metrics are mainly designed for evaluating word-level sequence, hence for character-level model, we would reconstruct the word-level sequence from the character-level prediction before the evaluation process. To demonstrate the benefits of the character-level representation in reducing potential OOVs, we futher evaluated each method by counting the number of <unk> tokens that appeared in all generated captions. 

\subsubsection{Results}

\input{table-auto-eval}

The overall performance of each model on the COCO-CN test set is  presented in Table \ref{tab:eval-cap}. As we can see from the table, our proposed
ABD-Cap model outperforms the baselines on all of the four caption evaluating metrics. Even more, the character-level models possess a zero OOV rate that is way less than the word-level models. The result justifies the superiority of the proposed character-level sentence generation for reducing OOVs. Therefore, the character-level ABD-Cap is deployed in the iCap system for the following real user study.

%% file: table-auto-eval.tex
\begin{table*}[tbh!]\centering
    \caption{\textbf{Performance of different models for VCSC without any user input}, \ie automated image captioning. \emph{Char} and \emph{Word} represent character-level and word-level decoders, respectively. Our proposed ABD-Cap model outperforms the baselines. Moreover, the character-level ABD-Cap has zero OOV rate.}
    \ra{1.3}
    \label{tab:eval-cap}
    \begin{tabular}{@{}|l|r|r|r|r|r|r|r|r|r|r|@{}}
   \hline
        & \multicolumn{2}{c|}{BLEU\_4} & \multicolumn{2}{c|}{METEOR} & \multicolumn{2}{c|}{ROUGE\_L} & \multicolumn{2}{c|}{CIDEr} & \multicolumn{2}{c|}{OOV rate (\%)} \\
    \hline
 \textbf{Model}  &  \emph{Char} & \emph{Word} & \emph{Char} & \emph{Word} & \emph{Char} & \emph{Word} & \emph{Char} & \emph{Word} & \emph{Char} & \emph{Word} \\
%\cmidrule{2-3} \cmidrule{4-5}
 \hline
 Show-and-Tell    & 24.0 & 25.5   & 27.1 & 27.4    & 47.8  & 48.4   & 70.6 & 72.4  & 0    & 3.8   \\
 \hline
 Show-and-Tell, backward decoder & 23.0 & 23.9   & 26.6 & 27.0   & 47.0  & 48.2    & 68.2  & 71.5 & 0    & 5.0   \\
 \hline
%        \midrule
 \emph{proposed ABD-Cap} & 24.4 & \textbf{26.0}     & \textbf{27.6} & 27.3     & 48.3   & \textbf{48.6}   & 71.7 & \textbf{73.7} & 0  & 7.4  \\
  \hline
%\bottomrule
    \end{tabular}
\end{table*}

%% file: eval-icap.tex
We first collect real-world interaction data from a user study. We then analyze the data in details to understand how users and the system interacted and to what extent the system assisted users to accomplish the annotation task.

\subsubsection{User Study}

To avoid any data bias towards COCO-CN, we constructed our annotation pool by randomly sampling images from MS-COCO with COCO-CN images excluded in advance.

Nineteen members in our lab, 14 males and 5 females, participated as volunteers in this user study. While mostly majored in computer science, the majority of the subjects have no specific knowledge about the iCap project. So they can be considered as average users for interactive image captioning. Our subjects performed the annotation task in their spare time. User actions in each session such as the editing history of the input box and the selection operation from the drop-down list were logged for subsequent analysis.

The user study lasted for one month, with 2,238 images  annotated and 2,238 sentences in total. Table \ref{tab:overview} summarizes main statistics of the gathered data. Depending on whether the system-suggested sentences were selected, user annotation  is divided into two modes, \ie \emph{fully manual} and \emph{interactive}. For 793 out of the 2,238 sentences (35.4\%), they were written in the interactive mode, \ie users selected the suggested sentences at least once. Moreover, the sentences written in the interactive mode tend to be longer and thus more descriptive than their fully manual counterparts. Also note the relatively smaller number of editing rounds in the interactive mode. These results encouragingly suggest that user-system interactions produce better annotations in a shorter time.

\input{table-overview}

%involved an annotation team comprised of 19 chinese native speakers. During each interactive captioning process, we implicitly recorded the annotation behavior of the involved human annotators and the corresponding reaction of the deployed image captioning model, which include edit history in the text entry box and selection operation from the dropdown list for the human annotators together with the provided completion candidates given current user input for the deployed captioning model. These recorded items provided us a solid basis for the subsequent analysis of the interactive process. 

\subsubsection{Analysis of an iCap Session}

We now analyze user-system interactions at the session level. The editing history of a specific user was recorded as a sequence of sentences $\{S_1,\ldots,S_T\}$, generated by scanning the input box every 0.2 second. If the text in the input changes between two consecutive scans, we consider an editing operation occur. To quantize the changes, we compute the Levenshtein distance\footnote{The Levenshtein distance between two Chinese sentences is the minimum number of single Chinese character edits (insertions, deletions or substitutions) required to change one sentence into the other} (LevD)~\cite{levd} between $S_i$ and $S_{i+1}$. On the basis of the pair-wise LevD, we derive the following two metrics:
\begin{itemize}
    \item \emph{Accumulated LevD}, computed over the sequence $\{S_1,\ldots,S_T\}$ by summing up all pair-wise LevD values. This metric reflects the overall amount of editing conducted in an iCap session. Smaller is better.
    \item $LevD(S,S_T)$ between an (incomplete) sentence $S$ and the final annotation. This metric estimates human workload. Specifically, $LevD(\varnothing, S_T)$ measures human workload required in the fully manual mode. Smaller is better. 
\end{itemize}

As Table \ref{tab:overview} shows, the sentence sequences in the interactive mode has an averaged \emph{Accumulated LevD} of 13.9, clearly smaller than that of the fully manual mode.

\begin{figure}[tbh!]
    \centering
    \includegraphics[width=0.95\columnwidth]{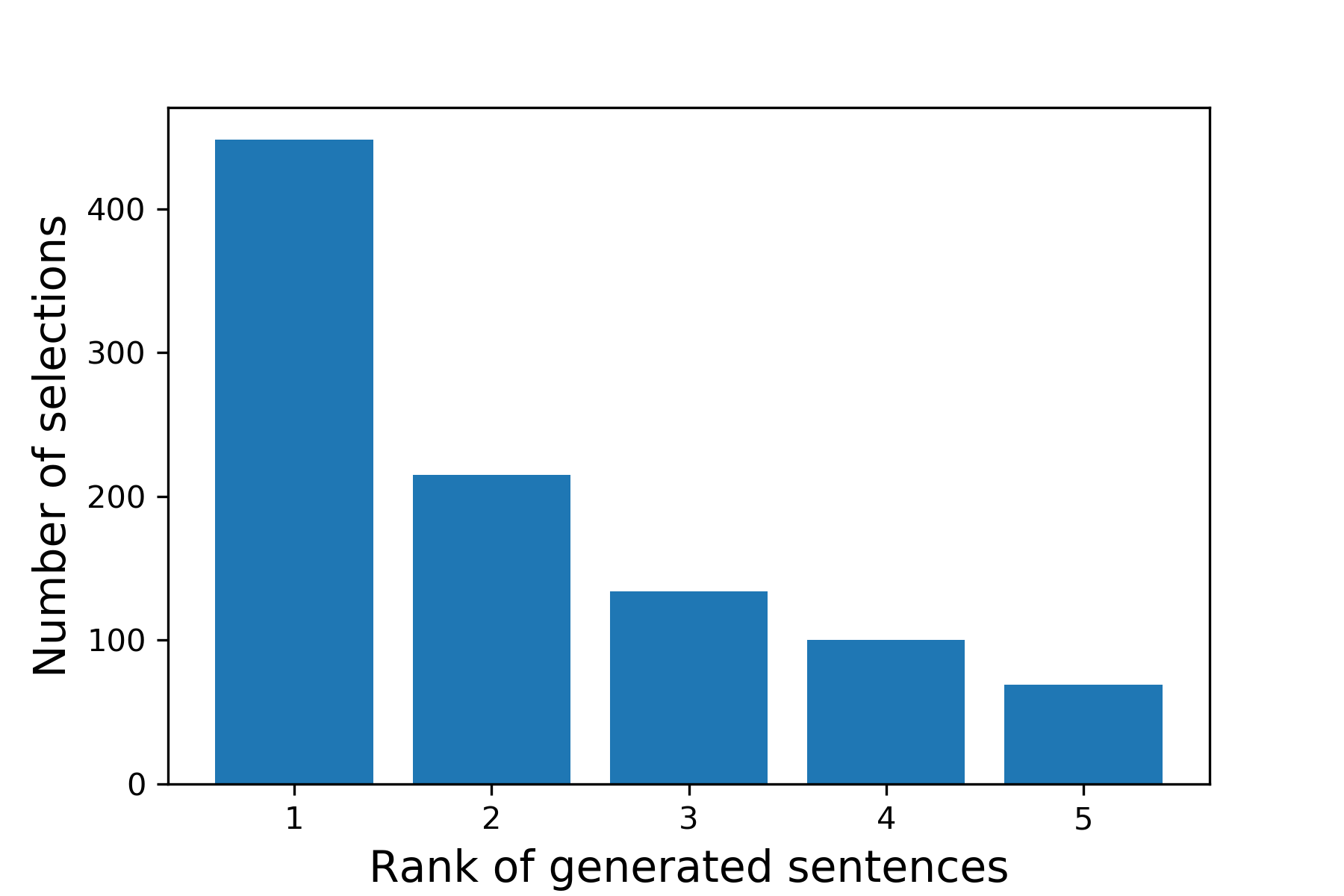}
    \caption{\textbf{Distribution of selection with respect to the ranks of generated sentences}. Sentences ranked at the top are more likely to be selected by users.}
    \label{fig:rank-count}
\end{figure}

\textbf{User behaviors}. For the 793 interactive annotations, a total number of 996 selections were recorded, meaning 1.2 selections per session on average. As shown in Fig. \ref{fig:rank-count}, more than half of the selections are made on the top-1 suggested sentences. Meanwhile, the other suggestions also get selected but with their chances decreasing along their ranks. The result demonstrates the effectiveness of beam search as well as the necessity of presenting multiple suggestions to the user.

%According to their ranking in the list of completion candidates, the distribution of the recorded selection operations are reported in Figure \ref{fig:rank-count}. As we can see, more than half of the selections are made on the top one ranked completion candidates (with selection ranking eqauls 1), and as the selection ranking increases from 1 to 5, the number of selections gradually decreases, which demonstrate the effectiveness of the beam search based ranking scheme.

\begin{figure}[tbh!]
    \centering
    \includegraphics[width=0.95\columnwidth]{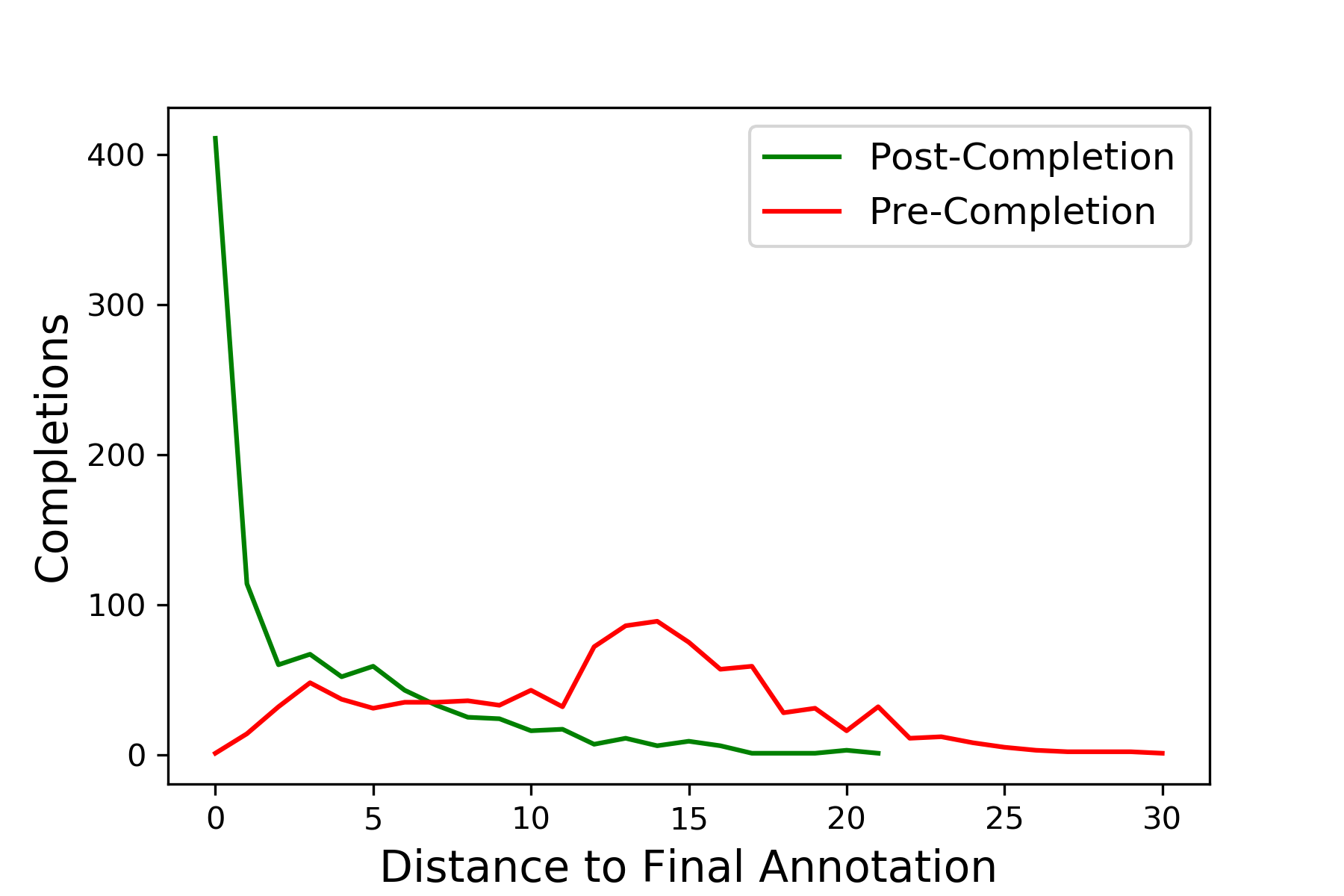}
    \caption{\textbf{Distribution of pre-completion / post-completion sentences's LevD to the final annotation}. Smaller LevD means post-completion sentences are more close to the final annotation.}
    \label{fig:completion}
\end{figure}

%In order to evaluate the role that the deployed image captioning model played in an authentic interactive captioning process, we zoom in on the 793 interactive annotation records in which the completion candidates generated by the captioning model have ever been selected by human annotators, and analyze the direct contribution (the 996 selected completions) made by the captioning model in these annotation sessions. 

% completions
\textbf{Influence of the VCSC model}. 
To study the direct influence of the VCSC model, let $S_s$ be the completed sentence selected by a given user at a specific editing round. By definition, the pre-completion sentence is obtained as $S_{s-1}$. We then calculate $LevD(S_{s-1},S_T)$ and $LevD(S_{s},S_T)$, which respectively measures the distance of the pre-completion and post-completion sentences to the final annotation.
The distribution curves of these two type of distances are shown in Fig. \ref{fig:completion}. The green curve which shows the distribution of $LevD(S_{s},S_T)$ is skewed towards the left side. The result clearly shows that the VCSC model helps the user input to converge to the final annotation, and thus reduces human workload.

\subsubsection{Comparing VCSC Models}

With the interaction data collected, we now perform a simulated experiment to compare two distinct instantiations of the VCSC module, namely ABD-Cap \emph{versus} Show-and-Tell.

Again, let $S_s$ be the candidate sentence selected by a given user at a specific editing round, and accordingly we have access to $S_{s-1}$ which is the user input of the VCSC module to generate $S_s$. Now, instead of ABD-Cap, we use Show-and-Tell to generate five candidate sentences. Comparing these candidates to the final annotation $S_T$ allows us to conclude which model provides better suggestions. 

%In previous stages, we have evaluated the performance of the proposed bidirectional captioning model and original unidirectional model for the conventional image captionig task, however we have not evaluated specifically the unidirectional and bidirectional completion algorithms that are based on these two kind of models. In order to justify the superiority of the proposed bidirectional completion algorithm over the original unidirectional completion algorithm, we could deploy them in authentic using scenarios and evaluate their performance by analyzing the selection times and selection rankings, but such evaluation requires the participation of a large number of users and can not be replicated. Therefore we propose to simulate the completion scenes (i.e. the presented image $I$ and partially completed user input $S$) and references (the finally submitted annotation) to evaluate the quality of the completion results by calculating Lev(C,A) as listed in Table \ref{tab:metrics}. Rather than artificially designing such simulated completion scenes, we exploited those authentic completion scenes within the 996 selected completions as mentioned in Section \ref{icap-analysis}, which could be a more reasonable choice to evaluate evaluate the proposed completion algorithms. 

\begin{table}[tbh!]
    \caption{\textbf{LevD between sentences produced by VCSC models to the final annotation}. Lower is better.}
    \label{tab:sim-perf}
    \begin{tabular}{lrrrrr}
    \toprule
     & \multicolumn{5}{c}{\textbf{Rank}} \\
     \cmidrule{2-6}
    \textbf{VCSC Model}     & 1    & 2    & 3   & 4  & 5  \\
    \midrule
    Show-and-Tell  & 5.69     & 5.93     & 6.05     & 6.21     & 6.43    \\ 
    ABD-Cap   & 4.99     & 5.17     & 5.28     & 5.56     & 5.76    \\ 
    \bottomrule
    \end{tabular}
\end{table}

\input{mid-case-table}

\textbf{Quantitative results}. Table \ref{tab:sim-perf} shows LevD between sentences generated by the two VCSC models to the final annotation. Lower LevD means the generated sentences are more close to the final annotation. The proposed ABD-Cap outperforms Show-and-Tell at all ranks. Moreover, the LevD increases as the rank goes up. This result confirms our previous finding (Fig. \ref{fig:rank-count}) that the sentences ranked at the top  describe images better and thus more likely to be selected by the user.

%For quantitative performance comparison between different completion algorithms, we follow the previous practice to calculate the Levenshtein distance between the generated completion candidates (ranking from 1 to 5) and the final submitted annotation as a evaluation score, by which the algorithm that achieves a lower score is deemed to have better performance. The overall performance of the two completion algorithms on each ranking are reported in Table \ref{tab:sim-perf}. As we can see from the table, the bidirectional completion algorithm outperforms the unidirectional completion algorithm to a large extent on all rankings and the performance within each method decreases as the ranking increases from 1 to 5, which is inline with our previous findings in the analysis of Figure \ref{fig:rank-count}.

\textbf{Qualitative results}. 
To better understand the differences between the two models for sentence completion, we present some typical results in Table \ref{tab:mid-completion-example}. In particular, the first two rows show cases where the user cursor is placed at the end of the current input text, \ie $S_{i,r}$ is empty. The last two rows show cases where the user cursor is placed in the middle. Compared to Show-and-Tell which
considers only text at the left-hand side of the cursor, our ABD-Cap model exploits texts at both sides, and thus suggest sentences more close to the final annotation.

%% file: table-overview.tex
\begin{table*}[tbh!]
    \caption{\textbf{Major statistics of the interaction data collected in our user study}. Smaller sequence length $T$, \emph{Accumulated edits}, and Levenshtein distance (\emph{LevD}) suggest less amount of human workload. }
    \label{tab:overview}
    \begin{tabular}{l r rr r rrrr}
    \toprule
                    & & \multicolumn{2}{c}{\textbf{Statistics per sentence}} && \multicolumn{4}{c}{\textbf{Statistics per iCap session}} \\
                \cmidrule{3-4} \cmidrule{6-9}
    \textbf{Annotation mode} & \textbf{Sentences} & \emph{Num. words} & \emph{Num. chars} & & \emph{Num. selections} & $T$  & \emph{Accumulated edits} & \emph{Accumulated LevD} \\
    \midrule
    \emph{Fully manual}          &  1,445    & 9.0  & 15.3  & & 0 & 11.6  &   10.6 &   18.9 \\   
    \emph{Interactive}    &  793    & 9.8  & 16.7  & & 1.2 &  9  & 8  & 13.9  \\
    \bottomrule
    \end{tabular}
\end{table*}

%% file: mid-case-table.tex
\begin{table*}[tb!]
\renewcommand{\arraystretch}{1}
\centering
\caption{\textbf{Some qualitative results showing how our iCap system responses to user input in a specific annotation session}. The cursor, marked out by $\land$, indicates the position where the user wants to edit. Compared to the baseline model (Show-and-Tell) which considers only text at the left-hand side of the cursor, our \emph{ABD-Cap} model exploits texts at both sides, and thus suggest sentences more close to final annotations.  Note that texts in parentheses are English translations, provided for non-Chinese readers.}
\label{tab:mid-completion-example}
\scalebox{0.84}{
\begin{tabular}{@{}c l l@{}}
\toprule
\textbf{Test image}  &  \textbf{Texts in an iCap session} \\
\midrule
\multirow{11}{*}{\includegraphics[width=4cm]{imgs/COCO_val2014_000000283254.jpg}}
& \emph{\textbf{User input at a specific moment}}: \\
& 一个$\land$ \\
& (A $\land$) \\
& \emph{\textbf{Top-1 sentence generated by a specific VCSC model}}: \\
& Baseline $\Rightarrow$  一个戴眼镜的男人正在打电话 \\
& (A man with glasses is on the phone) \\
& \emph{ABD-Cap} $\Rightarrow$  一个穿着西装的男人的黑白照片 \\
& (A man in a suit in a black-and-white photograph) \\

& \emph{\textbf{Final annotation}}: \\
& 一个穿着西装打着领带的男人坐在街道上的黑白照片\\
& (A man in a suit sit on the street in a black-and-white photograph) \\

\midrule

\multirow{10}{*}{\includegraphics[width=4cm]{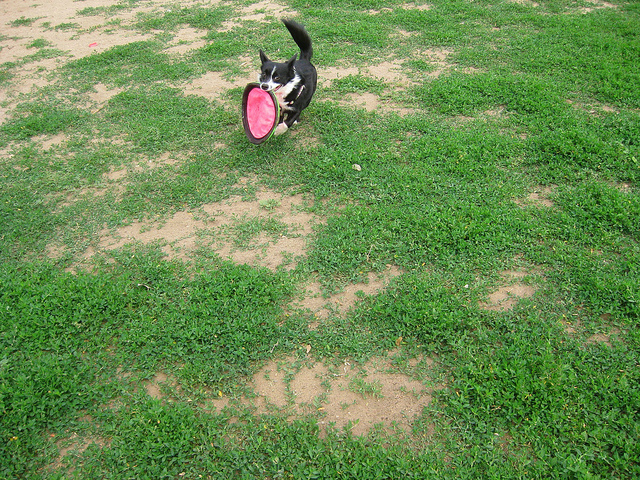}}
& \emph{\textbf{User input at a specific moment}}: \\
& 一只狗$\land$ \\
& (A dog $\land$) \\
& \emph{\textbf{Top-1 sentence generated by a specific VCSC model}}: \\
& Baseline $\Rightarrow$  一只狗在草地上奔跑 \\ 
& (A dog is running on the grass) \\
& \emph{ABD-Cap} $\Rightarrow$   一只狗在草地上玩飞盘 \\
& (A dog is playing frisbee on the grass) \\

& \emph{\textbf{Final annotation}}: \\
& 一只狗站在绿色的草地上玩飞盘\\
& (A dog is playing frisbee on the green grass)\\

\midrule

 \multirow{10}{*}{\includegraphics[width=4cm]{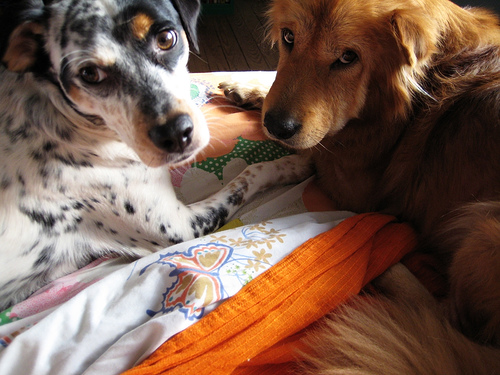}}
 & \emph{\textbf{User input at a specific moment}}: \\
 & 一只黑白相间的斑点狗和$\land$躺在沙发上 \\
 & (A black and white spotted dog and $\land$ lie on the sofa) \\
 & \emph{\textbf{Top-1 sentence generated by a specific VCSC model}}: \\
 & Baseline $\Rightarrow$  一只黑白相间的斑点狗和一只狗 \\ 
 & (A black and white spotted dog and a dog) \\
 & \emph{ABD-Cap} $\Rightarrow$   一只黑白相间的斑点狗和一只棕色的狗趴在沙发上 \\
 & (A black and white spotted dog and a brown dog lie on the sofa) \\
 & \emph{\textbf{Final annotation}}: \\
 & 一只黑白相间的斑点狗和一只棕色的狗趴在床上往回看向镜头\\
 & (A black and white spotted dog and a brown dog lie on the bed and look back at the camera)\\

\midrule

 \multirow{10}{*}{\includegraphics[width=4cm]{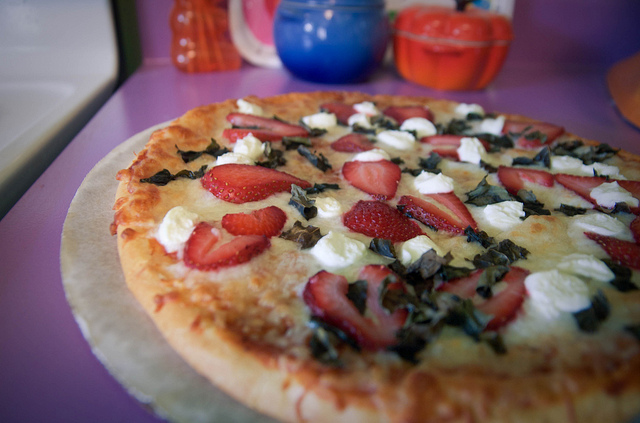}}
 & \emph{\textbf{User input at a specific moment}}: \\
 & 桌子上放着一个草莓$\land$披萨饼 \\
 & (On the table sits a strawberry $\land$ pizza) \\
 & \emph{\textbf{Top-1 sentence generated by a specific VCSC model}}: \\
 & Baseline $\Rightarrow$  桌子上放着一个草莓和一个比萨饼 \\
 & (On the table sits a strawberry and a pizza) \\
 & \emph{ABD-Cap} $\Rightarrow$   桌子上放着一个草莓和奶酪的比萨饼 \\
 & (On the table sits a strawberry and cheese pizza) \\
 & \emph{\textbf{Final annotation}}: \\ 
 & 桌子上放着一个草莓和奶酪的比萨饼\\
 & (On the table sits a strawberry and cheese pizza) \\

\bottomrule
\end{tabular}
}
\end{table*}